\documentclass{aa}
\usepackage{graphics}
\usepackage{epsfig}

\def\msol{M_\odot}

\def\te{T_{\rm eff}}

\def\simgr{\,\hbox{\hbox{$ > $}\kern -0.8em \lower 1.0ex\hbox{$\sim$}}\,}
\def\simle{\,\hbox{\hbox{$ < $}\kern -0.8em \lower 1.0ex\hbox{$\sim$}}\,}
\def\beq{\begin{equation}}
\def\eeq{\end{equation}}

\def\aj{AJ}                  
\def\araa{ARA\&A}             
\def\apj{ApJ}                 
\def\apjl{ApJL}                
\def\apjs{ApJS}               
\def\aap{A\&A}                
\def\mnras{MNRAS}             

\begin{document}


\title{Evolutionary models for low-mass stars and brown dwarfs: uncertainties
and limits at very young ages}
 \author{I. Baraffe\inst{1,2}, G. Chabrier\inst{1}, F. Allard\inst{1}
\and 
P.H. Hauschildt\inst{3}
}

\offprints{I. Baraffe}

\institute{C.R.A.L (UMR 5574 CNRS),
 Ecole Normale Sup\'erieure, 69364 Lyon
Cedex 07, France (ibaraffe, chabrier, fallard@ens-lyon.fr)
\and Max-Planck Institut f\"ur Astrophysik, Karl-Schwarzschildstr.1,
D-85748 Garching, Germany
\and
Center for Simulational Physics, University of Georgia
Athens, GA 30602-2451 (yeti@hobbes.physast.uga.edu)}

\date{Received /Accepted}

\titlerunning{Evolutionary models at very young ages}
\authorrunning{Baraffe et al.}
\abstract{
We analyse pre-Main Sequence evolutionary tracks for low mass stars
with masses $m \, \le \, 1.4 \, \msol$ based on the Baraffe et al. (1998)
input physics. We also extend the recent Chabrier et al. (2000) evolutionary
models based on
dusty atmosphere to young brown dwarfs down to one mass of Jupiter.
We analyse current theoretical uncertainties due to molecular line lists,
convection and initial conditions. Simple tests on initial conditions show
the high uncertainties of models at ages $\simle$ 1 Myr. We find a significant
sensitivity of atmosphere profiles to the treatment of convection at 
low gravity and $\te < 4000$ K, whereas it vanishes as gravity increases.
This effect adds another source of uncertainty on evolutionary
tracks at very early phases. We show that at low surface
gravity ($\log \, g \simle \, 3.5$,) the common picture of
vertical Hayashi lines with constant $\te$ is oversimplified. 
The effect of a variation of initial deuterium abundance is studied. 
We compare our models with 
evolutionary tracks available in the literature and discuss the main 
differences. We finally 
analyse to which extent current observations of young systems provide
a good test for pre-Main Sequence tracks.
\keywords{ stars: low-mass, 
brown dwarfs --- stars: evolution --- stars: Pre-Main Sequence }
}

\maketitle

\section{Introduction}

The development
of a new generation of stellar evolution
models based on the accurate coupling between interior
and atmosphere models yielded a major advance
in the description of very-low-mass stars (VLMS; Baraffe et al. 1995, 1997, 
1998, BCAH98) and 
substellar objects (brown dwarfs BD and giant planets GP; Burrows et al. 1997 
for objects with $\te < 2000$K;
Chabrier et al. 2000a, CBAH00). 
One of the main advantages  of such models over previous generation models is the direct comparison
between theory and observation in  colour-colour and colour-magnitude diagrams (CMD).
Several observational tests now assess the validity of the theory devoted
to  the description of low-mass ($\simle 1 \msol$) astrophysical objects   
(see Allard et al. 1997; Chabrier and Baraffe 2000, for recent reviews).
Although some discrepancies
between models and observations still remain, uncertainties due
to the input physics are now significantly reduced.  

In previous papers, we have developed evolutionary models of VLMS and BDs
(BCAH98; CBAH00),
taking into account the most recent improvements of the physics
describing
the interior  (equation
of state for dense plasmas, screening factors) and the atmosphere 
(molecular opacities, formation of dust) devoted to the analysis of
objects with an age $t \simgr$ 100 Myr.
Comparison between observations and models for very
young objects ($t <$ 100 Myr) are 
more uncertain, in particular from  the observational viewpoint, 
mainly for two reasons :
(1) extinction due to the surrounding dust
modifies both the intrinsic magnitude and the colours of the objects and
 (2)
spectra of
 very young objects ($t \simle$ 1 Myr) may still be affected by the presence
of an accretion disk or circumstellar material residual from the protostellar 
stage.
On the theoretical side, an important source of uncertainty at the
very early stages of evolution is the choice of the initial conditions,
resulting from prior protostellar collapse and accretion phases. The simple
 picture of non-accreting objects contracting from large initial
radii, as used in our models, is clearly an idealized description of reality
  (Stahler, Shu \& Taam 1980;
Stahler 1983, 1988; Hartmann, Cassen \& Kenyon 1997 and references
therein).  

In spite of these uncertainties, numerous surveys 
are devoted to very young clusters because of the potential
detection of substellar objects down
to planetary masses  (Zapatero et al. 2000; Lucas et al. 2001),
and the determination of the very (sub)stellar initial mass functions (IMF), free of
dynamical evolution effects that affect older clusters.
Consequently, a wealth of data for low mass objects with ages spanning 
$\sim$ 1-10 Myr is available and provides the basis for a better
understanding of the early phases of stellar evolution.
Given the reliability of the present theory for VLMS and BD, a confrontation
of the models with the observational data of such very young objects provides important insight
into the early epoch of protostellar collapse and accretion phases.
For this purpose, however, a correct appreciation of the model uncertainties
and limitations is required. 
 
For this reason, this paper is specifically devoted to the analysis
of evolutionary 
models by BCAH98 and CBAH00 at
early phases of evolution ($t \simle$ 100 Myr). 
The input physics is briefly recalled
in section 2, followed by a discussion of uncertainties specific to
evolutionary tracks at early ages
(\S 3). Our models are compared to other
evolutionary tracks available in the literature in \S 4
and the confrontation to observations  is presented in  \S 5.    

\section{Model description}

\subsection{Evolutionary tracks}

The models analysed in the present paper
are based on the input physics already described in 
BCAH98 and CBAH00. 
Both sets of models use the same ingredients
describing the stellar interior but use different sets of atmosphere models,
which provide the outer boundary conditions and the synthetic spectra.
The BCAH98 evolutionary tracks are based on the
non-grey atmosphere models by Hauschildt, Allard and Baron (1999a). 
These models are dust-free and are appropriate
to the description of objects with effective temperatures
$\te\,  \simgr \, 2300$ K. The CBAH00 models are based on atmospheres including
the formation and opacity of dust (Allard et al. 2001, hereafter DUSTY models
). As illustrated
in CBAH00, dust must be taken into account in order to explain the near-IR
colors of late M-dwarfs and L-dwarfs. The latter models
are thus more appropriate to
the description of objects with $\te \, \simle \, 2300$ K. 
As emphasized in CBAH00, the DUSTY models are not appropriate for the
description of spectral and photometric properties of methane dwarfs
($\te <$ 1600K), which require a different treatment of dust (the so-called
COND models in CBAH00 and Allard et al. 2001). 
Evolutionary models for such cool objects
will be presented elsewhere (Chabrier et al. 2001, in preparation).

\begin{figure*}
\psfig{file=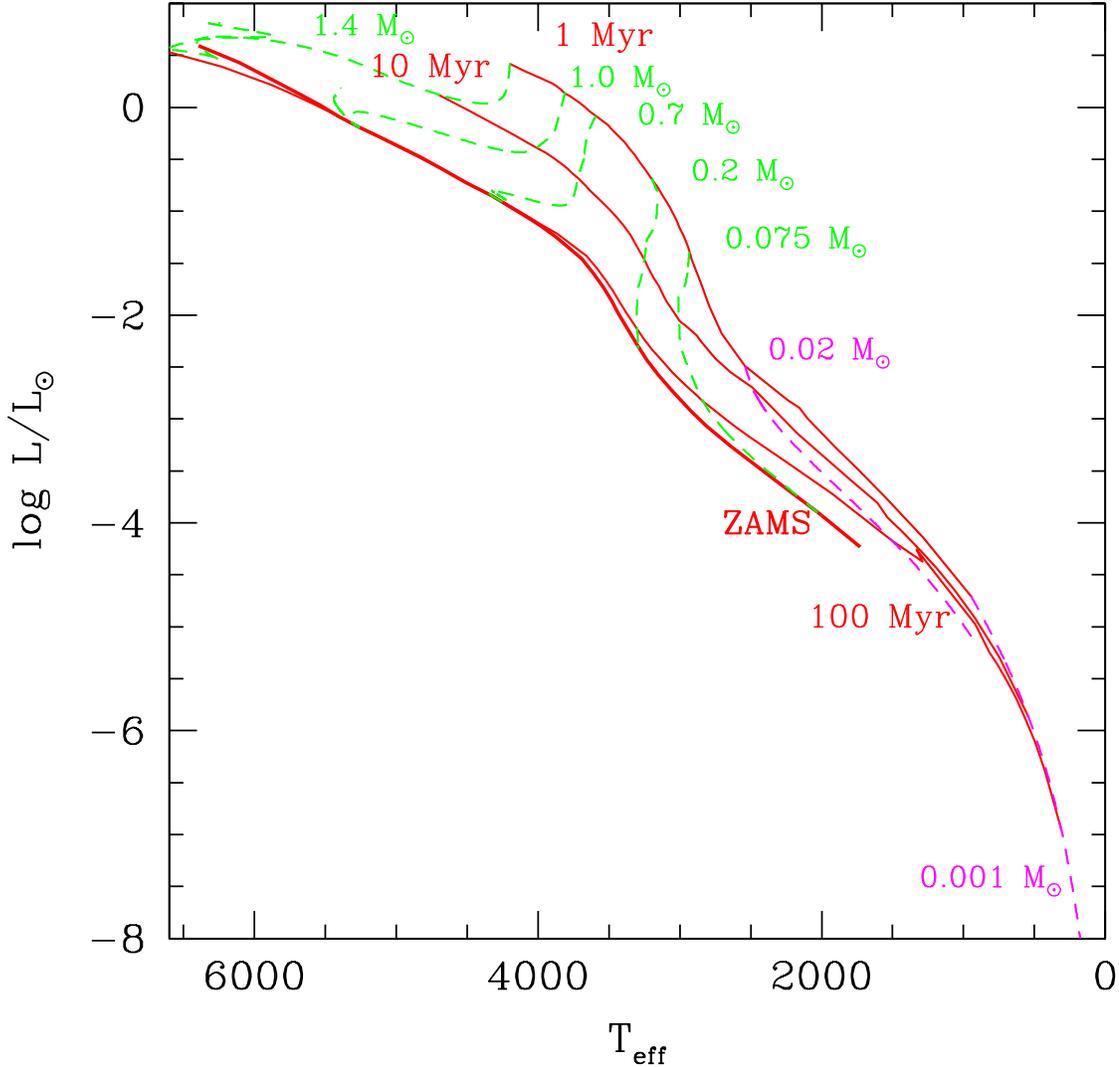,height=160mm,width=160mm} 
\caption{Evolutionary tracks in the Hertzsprung-Russell diagram for masses
from 1.4 $\msol$ to 0.001 $\msol$ (dashed lines) 
and ages spanning from 1 Myr to the ZAMS
(for stars). Several isochrones for 1, 10 and 100 Myr
are indicated by solid lines from right to left. The location of the ZAMS
for stars down to 0.075 $\msol$ is also indicated (left solid line).}
\label{fig1}
\end{figure*}

\begin{figure}
\psfig{file=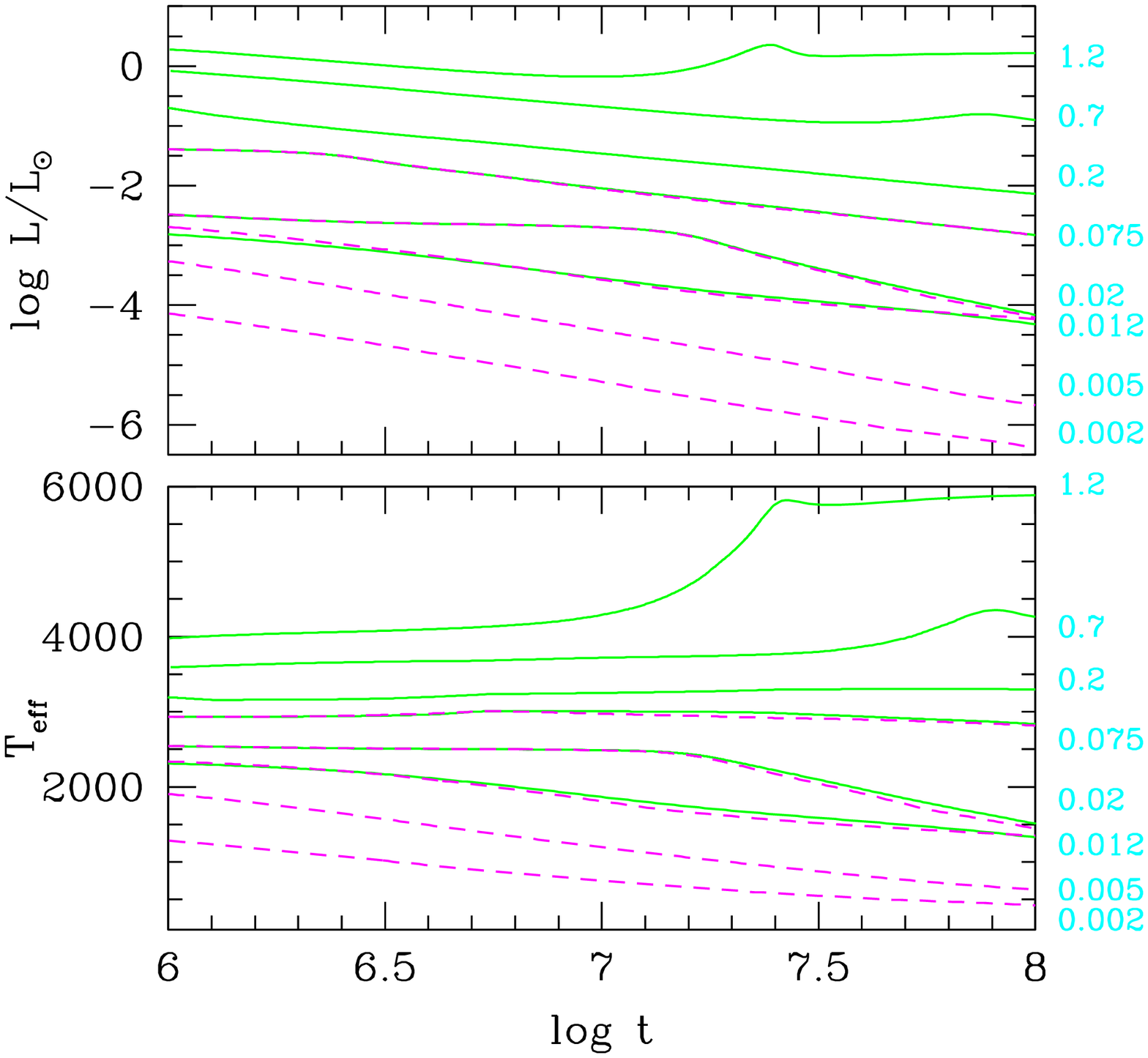,height=110mm,width=88mm} 
\caption{Evolution of luminosity and effective temperature
as a function of time (in yr) for masses
from 1.2 $\msol$ to 0.002 $\msol$. Masses (in $\msol$)
are indicated on the right hand side of the figure. The solid lines are the
BCAH98 models. The dashed lines are dusty-models based on the CBAH00 input
physics.}
\label{fig2}
\end{figure}

The BCAH98 grid covered a mass range from 0.02 $\msol$ to 1.4 $\msol$,
for ages $\ge$ 1 Myr up to the Main Sequence for stars.
Originally, the CBAH00 grid covered masses from 0.01 $\msol$ to 0.1 $\msol$
for ages $\ge$ 10 Myr.
In the present work, we extend it down to 1 $M_{\rm J}$ (10$^{-3} \msol$)
and ages $\ge$ 1 Myr.
Figure \ref{fig1} presents the complete
grid of models in a Hertzsprung-Russell diagram
(HRD) from 0.001 $\msol$ to 1.4 $\msol$. The small variation of radius 
with mass and age in the substellar regime (see e.g. Chabrier \& Baraffe, 2000) yields the merging towards very similar tracks below the 
hydrogen-burning limit.
Figure \ref{fig2} displays the
 time evolution of the effective temperature and luminosity 
for selected masses.   
Objects below 2 $M_{\rm J}$
evolve essentially with $\te <$ 1600K (see Figs. \ref{fig1} and 2),
even at very early ages. Their
 atmospheric properties are thus better described by the COND models.  

The initial conditions of the models are described in the next section.
The stellar/substellar transition is located
at $m\sim$ 0.075 $\msol$. Below this limit, objects become partially degenerate
 and their nuclear energy production  
cannot compensate the energy lost by radiation (e.g $L_{\rm nuc} < L$),
which is required to reach the Main Sequence.
The deuterium  burning minimum mass is $m\sim$ 0.012 $\msol$ 
(Saumon et al. 1996; Chabrier et al. 2000b). 
The initial D-burning phase lasts less than
1 Myr for $m \simgr 0.2 \msol$, between 1 and 5 Myr for  $0.05 \simle m \simle 0.2 \msol$
and almost 20 Myr for a 0.02 $\msol$ brown dwarf (see Chabrier et al. 2000b,
 for details). 

As shown in CBAH00, grains have little effects on the evolution of $L(t)$ and $\te(t)$,
 because of the reduced dependence of evolution upon opacity, 
 $L(t)\propto \kappa_R^{\sim 1/3}$,
$\te(t) \propto \kappa_R^{\sim 1/10}$ (Burrows \& Liebert, 1993).
We verified that the difference between the different TiO and H$_2$O molecular 
line lists used in BCAH98 and CBAH00 models (see \S 3 in CBAH00), respectively, affect essentially the outer atmospheric
layers, and thus the synthetic spectra
and colors, but not the deeper
atmospheric layers, and thus  the outer boundary conditions.
The effect of these different molecular linelists on the evolution of the effective
temperature
$\te(t)$ and the bolometric luminosity $L(t)$ is small, less than 100 K in
$\te$ and 10\% in $L$  at a given age. This is illustrated in Figure  2,
where tracks based on the BCAH98 input physics (solid lines) are compared
to the CBAH00 tracks (dashed lines) for the same masses (0.075, 0.02
and 0.012 $\msol$). 
As stressed in CBAH00 and
Baraffe et al. 2001, the computation 
of more reliable H$_2$O and, to a lesser extend, TiO linelists is badly needed to solve this shortcoming in the present theory.

\section{Main uncertainties: initial model and convection}

\subsection{Initial conditions}

As mentioned in the previous section, although shortcomings still remain
in current molecular opacities, the resulting uncertainty on
 the evolution is small. One of the main source 
of uncertainty for models at early stages of evolution is the choice of the initial
conditions. Most of low mass pre-Main Sequence (PMS) models available in the 
literature (D'Antona \& Mazzitelli
1994, 1997; Burrows et al. 1997; BCAH98; Siess et al. 2000) start from
arbitrary initial conditions, totally independent on the outcome of the prior proto-stellar
collapse and accretion phases. The initial configuration is
that of a fully convective object
 starting its contraction along the Hayashi line
from arbitrary large radii. Evolution starts prior to or at
central deuterium ignition,  with
initial central temperature  $\sim \, 5 \, 10^5 $ K. 
According to studies of
low-mass protostellar collapse and accretion phases, such initial conditions
are oversimplified, and low mass objects should rather 
form with relatively small
radii (Hartmann et al. 1997, and references therein). 
Based on spherical accretion protostellar models, Stahler (1983, 1988)
defined a birthline in the Hertzsprung-Russell diagram where
young objects become visible. Evolutionary tracks should
then start from this birthline, which fixes the age $t$ = 0.
Ages determined from models based
on the above-mentioned oversimplified conditions
should then be corrected accordingly, with substantial corrections
for systems younger than a few Myr (see Palla and Stahler 1999).
Furthermore, collapse and accretion are unlikely to proceed spherically.
Spherical collapse does not consider angular momentum transport, an important issue
of the early phases, which affects the subsequent cooling and formation of the protostar.
Hartmann et al. (1997) recently illustrate the sensitivity of
the birthline locus assuming that accretion proceeds through a disk
rather than spherically. This work stresses again the high uncertainty of
assigning ages from HRD positions for the youngest objects.

This analysis demonstrates convincingly that assigning an age to objects younger than a few Myr is totally meaningless
when the age is based on models using oversimplified initial conditions.
As shown in the next section,
for  $t \, \simle \, 1$ Myr,
the evolutionary tracks themselves are sensitive to the initial conditions,
whereas after a few Myr, the models converge toward the same
track. This is the main reason why we provide confidently evolutionary models
for ages $t \, \ge \, 1$ Myr, considering
that below such ages, models are too uncertain. To solve this substantial uncertainty requires
 the consistent evolution between
the 3D collapse of the protostellar phase and the subsequent PMS evolution.
 
\subsubsection{Effect of the initial radius and the mixing length parameter}
 
In this section we examine the effect of the variation of the
initial radius (or initial gravity) in the initial conditions. 
For this purpose we have extended our grid 
of atmosphere models to gravities $\log g \, < \,  3.5$. 
We have checked the validity of the plane-parallel approximation used in these models
by comparison with similar
atmosphere models which take into account the effects of spherical geometry.
These effects are found to become important only for surface gravities 
$\log g \, \simle \,  2$ (Hauschildt et al. 1999b; 
Allard, Hauschildt \&  Schweitzer 2000),
well below the range of gravities characteristic of
the evolution of young low mass objects. 

We first calculate a set of models, labeled (A), 
with initial radii fixed to obtain
 initial 
surface gravities $\log g \, \sim \,  3-3.5$ and initial thermal
 time-scales $t_{th} \sim$ a few Myr.
For masses above the 
deuterium burning minimum mass (DBMM), $m \, \simgr \, 0.02 \,
\msol$,  such tracks start at deuterium 
ignition, with central temperature
 $\sim 5 \, 10^5 \, - \, 10^6$K, 
defining a Zero-Age-Deuterium Burning Sequence.
Such initial conditions are similar to that used in BCAH98 and CBAH00.
A second set of models, labeled (B), starts with larger radii such that 
the
 initial surface gravity $\log g \, \sim \,  2.5$. These initial models
are more luminous than the previous ones,  with central 
temperatures  below 5 10$^5$K and initial thermal
 time-scales $t_{th} \sim \, 10^5$ yr.

We also analyse the sensitivity of the models to the mixing length
$l_{\rm mix}$,  characteristic of the mixing length
formalism (MLT) used to describe convection.
We used two different values of the mixing length parameter
$\alpha_{\rm mix}$\footnote 
{$\alpha_{\rm mix} = l_{\rm mix}/H_{\rm P}$, with $H_{\rm P}$ 
the pressure scale-height} = 1 and 2 
(see  \S 3.2 for justifications).

\begin{figure}
\psfig{file=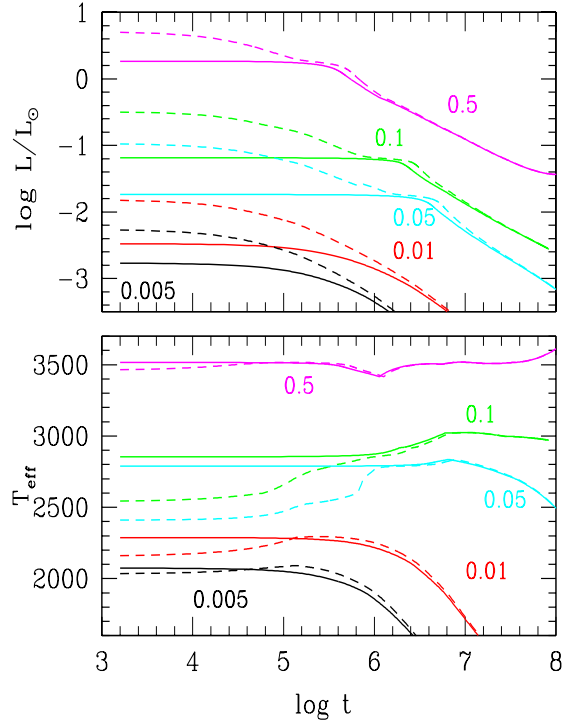,height=110mm,width=88mm} 
\caption{Effect of the initial radius on the
evolution of luminosity and effective temperature
as a function of time (in yr) for several masses (indicated near the curves
in $\msol$). The solid lines correspond to the first set of models (A)
with initial gravity $\log g \, \sim \,  3-3.5$  and the dashed lines 
correspond to models (B) starting with initial gravity 
$\log g \, \sim \, 2.5$. The mixing length is $l_{\rm mix} = H_{\rm P}$
}
\label{fig3}
\end{figure}

\begin{figure}
\psfig{file=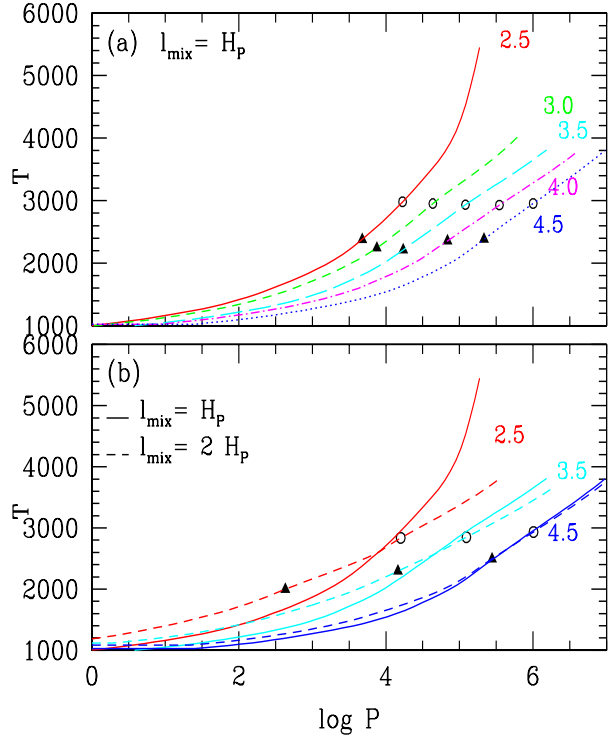,height=110mm,width=88mm} 
\caption{{\bf (a)} Gravity effect on $P$ - $T$ 
atmosphere profiles with $\te = 2500$ K and $l_{\rm mix} = H_{\rm P}$.
Pressure is in dyne.cm$^{-2}$ and temperature  in K.
The surface gravity varies from $\log g$ = 2.5
to 4.5, as indicated on the figure. The open circles on each curve correspond
to an optical depth $\tau$ = 1. 
The filled triangles indicate the onset of convection. 
The inner atmosphere profiles stop at $\tau$ = 100, where the outer boundary 
condition for the inner structure is defined.
{\bf (b)} Effect of a variation of $l_{\rm mix}$ for atmosphere
profiles with $\te = 2500$ K and gravity $\log g$ = 2.5, 3.5 and 4.5.
Symbols on the dashed curve are the same as in (a).
}
\label{fig4}
\end{figure}

The time evolution of $L$ and $\te$ for models (A) (solid lines) and (B) 
(dashed lines)
is displayed in Figure \ref{fig3} for several masses and for $\alpha_{\rm mix}$
 = 1. 
Figure \ref{fig3} shows the importance of the initial radius
on $\te$, and thus on $L$, during the first Myr of evolution. After a few Myr, however, the tracks
based on different initial radii merge, for a given mass. 
For masses ranging from $\sim$ 0.01 to 0.5 $\msol$, the two sets of models start at very different $\te$.
Models starting
with the lowest gravity are cooler by up to several hundreds K compared
to initially denser, less luminous models with the same mass.
Note also that for models (B), $\te$ increases during the early evolution,
under contraction, 
in contrast to the first set of models with initial $\log g \, \simgr \,  3.0$.
This is the consequence of the different surface gravities, which strongly
affect the atmosphere profiles for $\te$ = 2200-3500K,
as illustrated in Figure \ref{fig4}.  As shown in this figure,
{\it for this range of effective temperatures}, the atmosphere profiles also show
an extreme sensitivity to the mixing length $l_{\rm mix}$ at low gravity 
($\log g \, < \,  3.5$)
 but become rather insensitive to
a variation of  this parameter for
$\log g \,\ge   \,  3.5$ (see  Fig. \ref{fig4}b).  
 
\begin{figure}
\psfig{file=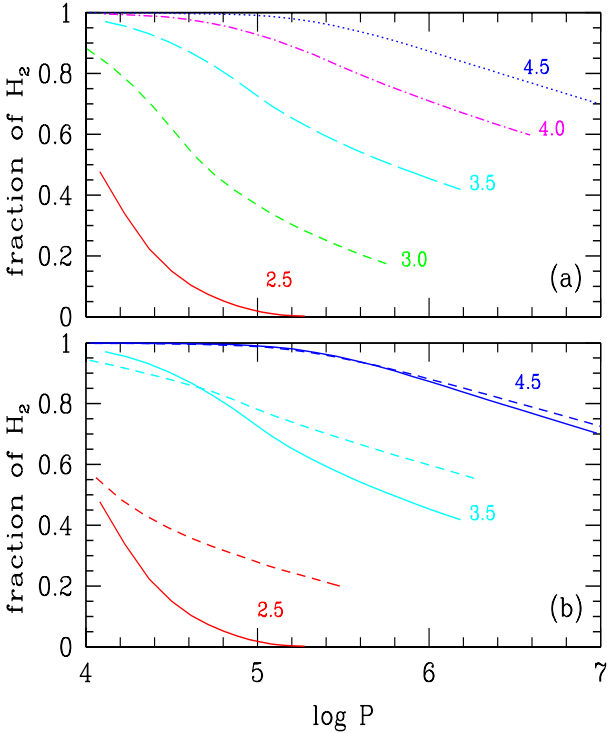,height=110mm,width=88mm} 
\caption{{\bf (a)} Number fraction of molecular H$_2$ as a function of $P$ (in dyne.cm$^{-2}$)
along the profiles displayed in Fig. \ref{fig4} with   
$\te = 2500$ K, $l_{\rm mix} = H_{\rm P}$ and
 different gravities $\log g$ (indicated near the curves).
 {\bf (b)} Same as in (a)
for  $l_{\rm mix} = H_{\rm P}$ (solid lines) and $l_{\rm mix} = 2 H_{\rm P}$
(dashed lines). 
}
\label{fig5}
\end{figure}

Such a sensitivity of the atmospheric profiles to gravity and
mixing length in this range of $\te$
stems from the less efficient formation of molecular H$_2$ as gravity decreases.  
Figure \ref{fig5} displays the fraction of H$_2$ along
the atmosphere profiles shown in Fig. \ref{fig4} (below the photosphere).
 For $\te$ = 2500K and $\log g = 2.5$,
the fraction $n_{\rm H_2}$  of H$_2$   decreases rapidly along the inner
profile. For $\alpha_{\rm mix}$ = 1, it
becomes negligible at optical depth $\tau$ = 100, 
where the boundary condition for the inner structure is 
defined (see Chabrier \& Baraffe
1997, their \S 2.5). 
For $\log g \, \ge \,  3.0$ (see Fig. \ref{fig5}), 
 $n_{\rm H_2}$ exceeds 20\%, even in the deepest atmospheric layers. 

As described in Chabrier \& Baraffe 
(2000, \S 2.2.2),  hydrogen atom recombination yields a rapid increase of
the H$_2$ Collision Induced Absorption opacity
($\kappa_{\rm CIA} \propto \, n_{\rm H_2}^2$) and a
 decrease of the adiabatic gradient . 
Both effects favor convection, yielding
a flatter atmosphere profile, {\it i.e} a decrease of $T$  at fixed $P$. 
This in turn favors the formation of H$_2$ and increases $n_{\rm H_2}$
at fixed $P$, increasing again the opacity and the efficiency
of convection.  
This illustrates the non-linear response of the atmospheric structure
to the formation
of H$_2$, as soon as its fraction becomes significant.
It explains the huge effect of gravity when this latter increases from 
$\log \, g\,=$ 2.5 to 3.0 on the atmospheric profile, as illustrated in Figure \ref{fig4}a. 
The decrease of $\nabla_{ad}$ in regions of H$_2$
formation favors adiabatic convection, limiting the extension of
super-adiabatic layers and thus the sensitivity of
the thermal profile to the mixing length parameter. 
This explains the strong dependence of the atmosphere profile on $\alpha_{\rm mix}$ for
$\log \, g$ = 2.5 (see Fig. \ref{fig4}b).
At higher gravities convection described
by the MLT is essentially adiabatic and almost insensitive
to the choice of $\alpha_{\rm mix}$ (see \S 3.2).

A similar behavior is found in the range of effective temperatures
2200 K $\simle \, \te \, \simle$ 4000K. 
Above $\sim$ 4000K, the 
structure is too hot for H$_2$ to form in significant fraction whereas
below  $\te \, \sim \, 2200K$,
the outer layers are dense and cool enough for H$_2$ to form efficiently, even at 
$\log \, g \, = \, 2.5$, .

The drastic modification of the atmospheric profile,
for $\alpha_{\rm mix}$ = 1, as gravity increases
from $\log g$ = 2.5 to $\log g \, \ge$ 3.0 explains the different $\te$'s 
for models  starting from different initial radii 
(Fig. \ref{fig3}). 
For models (B), starting with $\log g$ = 2.5, gravity increases as
contraction proceeds and favors H$_2$ molecular formation, 
yielding significantly flatter atmospheric profiles, as mentioned above.
This yields
a significant increase of $\te$ at fixed mass $m$ as contraction proceeds (Chabrier \& Baraffe, 
2000, \S 3.2). 
For $\alpha_{\rm mix}$ = 2, the sensitivity of atmosphere profiles
to $g$ is less pronounced, but still yields up to 200 K differences
between models (A) and (B) at a given age.   

\begin{figure*}
\psfig{file=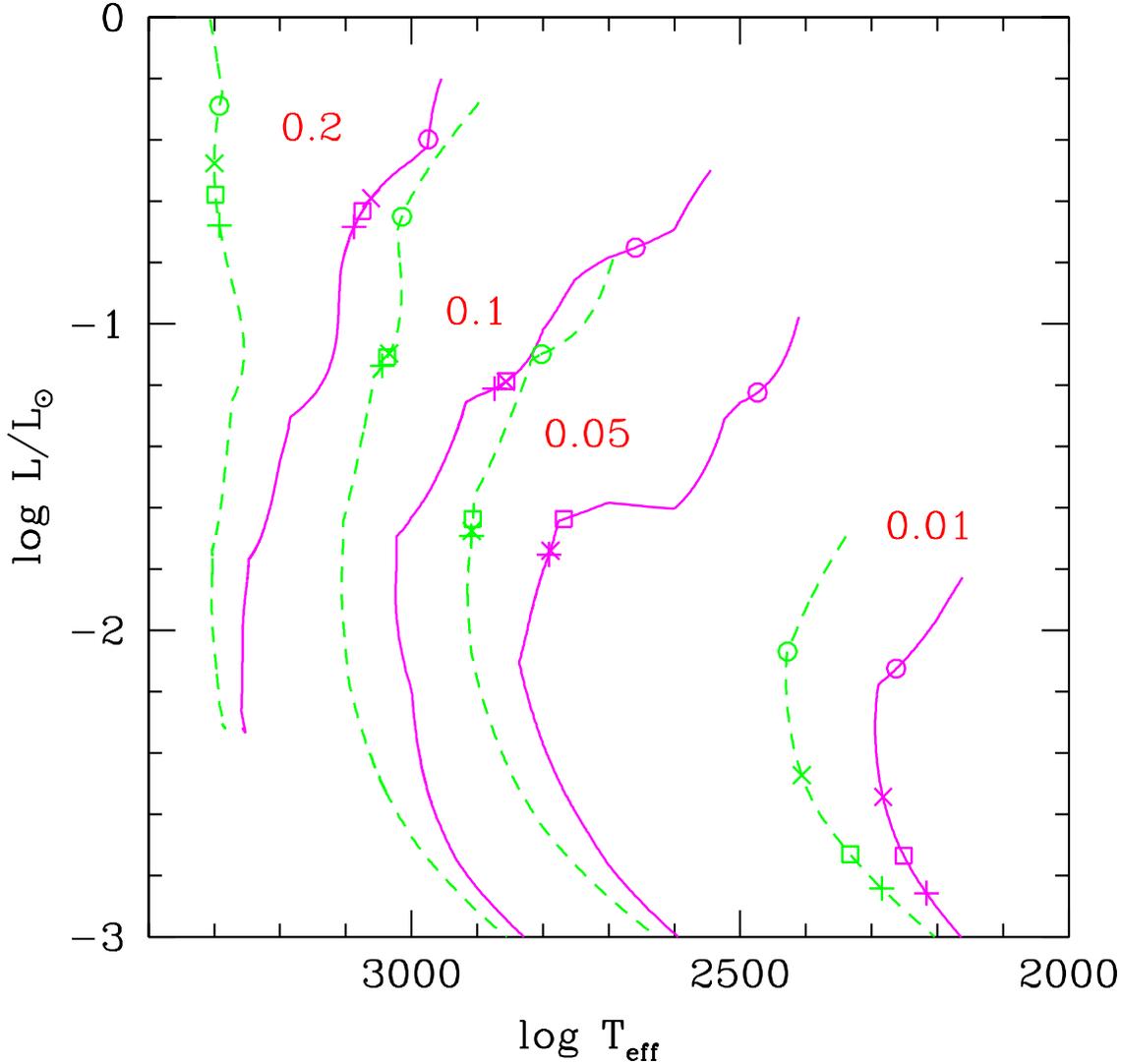,height=160mm,width=160mm} 
\caption{Evolutionary tracks in the Hertzsprung-Russell diagram for masses
from 0.2 $\msol$ to 0.01 $\msol$, as indicated. The solid curves correspond 
to $\alpha_{\rm mix}$ = 1 and the dashed curves to $\alpha_{\rm mix}$ = 2.
Ages of 0.1 and 1 Myr are indicated by respectively crosses and plus for 
models (A) with initial gravity $\log g$ = 3.0-3.5 
and by open circles
and squares for models (B) with initial $\log g$ = 2.5. Note that for models
(A) the initial position (age=0) is essentially the same as the position 
at 0.1 Myr (cross) and can differ significantly from position of models (B)
at $t=0$.  
}
\label{fig6}
\end{figure*}

The effect on evolutionary tracks in a HRD is displayed in Figure \ref{fig6}.
Note that evolution along the Hayashi line does not necessarily
proceed at constant $\te$, because of the effects described above.
The common picture of vertical (constant $\te$) Hayashi tracks is therefore an oversimplified
picture of PMS evolution.
As demonstrated in
 the pioneering papers of Hayashi (1961) and Hayashi \& Nakano (1963),
fully convective and adiabatic
objects which have H$_2$ dissociation zones contract almost
vertically in the HRD. The very low value of $\nabla_{ad}$ ($< 0.1$)
in such zones is responsible for such evolution at constant $\te$.
Taking into account super-adiabaticity in convective
layers displace the Hayashi line toward lower effective temperature.
This is equivalent to a decrease of $\alpha_{\rm mix}$ in our calculations.
Hayashi \& Nakano (1963) have also shown that omitting the presence of
H$_2$ molecules yields an evolution proceeding at
 decreasing $\te$ (from the left
to the right in a HRD), rather than at constant $\te$. 
Although based on a simplified  treatment of the atmospheric properties,
opacities and
equation of state, these basic works already illustrated the 
extreme sensitivity of the shape of PMS tracks for low mass objects 
to super-adiabaticity {\it and} to the degree of H$_2$ formation/dissociation.

Models (A) and (B) follow {\it the same track} for a given mass and
$\alpha_{\rm mix}$, but do not reach {\it the same position at the same age}.
Significant differences appear at ages $\simle$ 1 Myr but vanish after
a few Myr. We thus consider 1 Myr as the 
characteristic time required to forget our
arbitrary initial conditions and below which models are too 
sensitive to input physics and thus too uncertain. 

\subsubsection{Effect of the initial deuterium abundance}

Deuterium burning plays a key role  during
the protostellar collapse phase (Stahler 1988) and the following
$\sim$ 10 Myr of evolution. 
In a previous paper, we focused on the initial 
deuterium burning phase (Chabrier et al. 2000b), adopting an initial
mass fraction [$^2D_0$] = 2 10$^{-5}$,
characteristic of the local interstellar medium (hereafter LISM, Linsky 1998).  
In the present section we examine the effect of a variation of [$^2D_0$]
by a factor of 2 on the early stages of evolution. 
Such a variation is motivated by a recent deuterium abundance determination
 towards quasars, which  suggests a primordial abundance 
 only slightly larger (by less than a factor of 2) than
the LISM value (O'Meara et al. 2001). 

\begin{figure}
\psfig{file=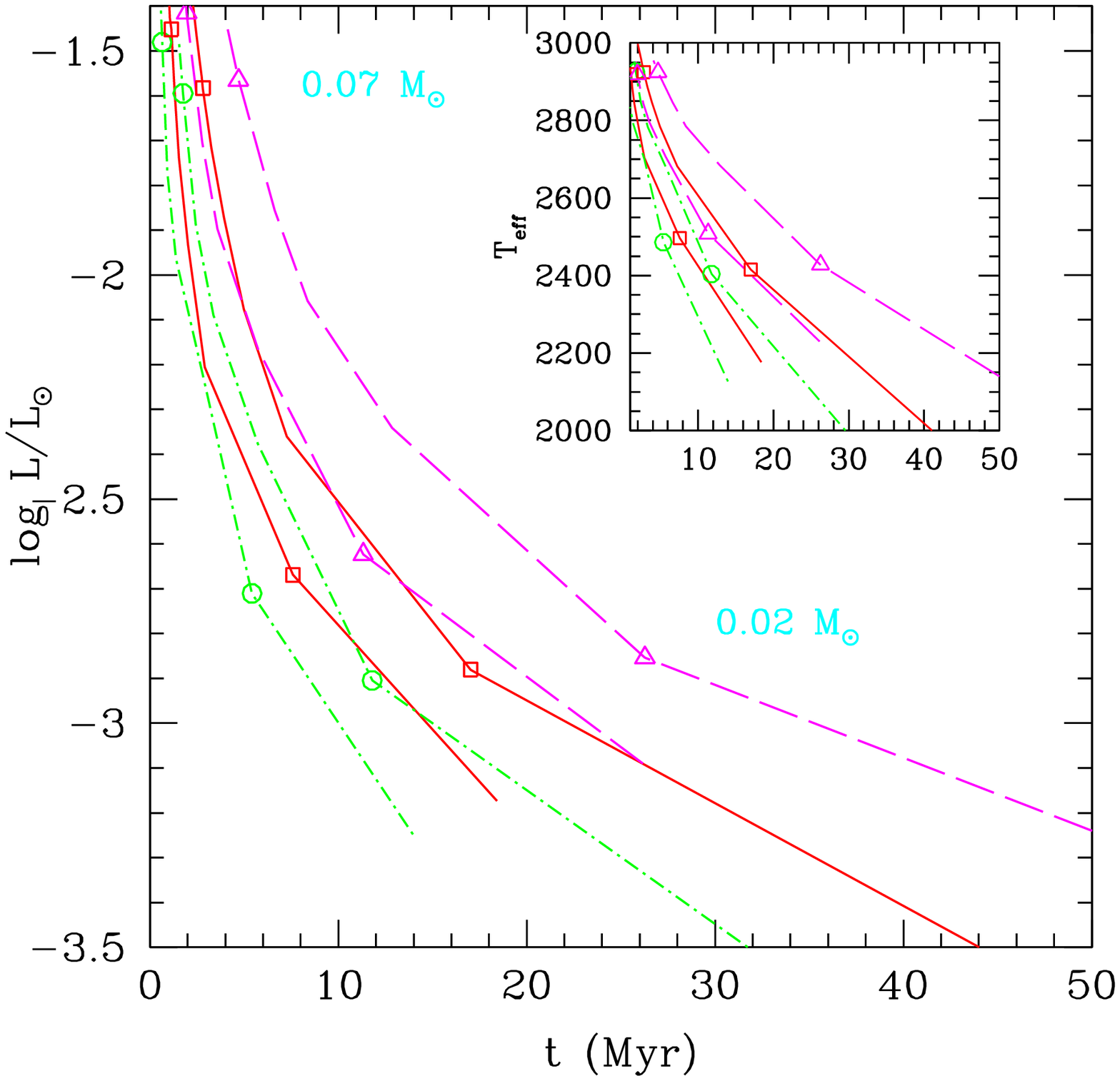,height=110mm,width=88mm} 
\caption{Deuterium depletion curves as a function of age in $L$
and $\te$ for different initial D abundances: [$^2D_0$] = 10$^{-5}$ 
(dash-dotted curves), [$^2D_0$] = 2$\times$10$^{-5}$ (solid curves) and
[$^2D_0$] = 4$\times$10$^{-5}$ (dashed curve). For each initial value of
 [$^2D_0$], two curves are shown corresponding to
the 50\% (left curve) and 99\% (right curve) depletion limits.
The symbols on the curves indicate the location of respectively
0.07 $\msol$ and 0.02 $\msol$ brown dwarfs. 
}
\label{fig7}
\end{figure}

The main effect of a variation of the initial [$^2D_0$] is the modification
of the age at a given $L$ and $\te$, with the most important effect
for masses $m \, \simle 0.07 \, \msol$ (Figure \ref{fig7}).
For a standard [$^2D_0$], a 0.07 $\msol$ depletes by a factor of 
100 its initial deuterium content within $\sim 3$ Myr  and
a 0.02 $\msol$ needs $\sim 17$ Myr (see Chabrier et al. 2000b).
If [$^2D_0$] is increased by a factor of 2, the 0.07 $\msol$ brown dwarf
then needs $\sim 5$ Myr and the 0.02 $\msol$ requires $\sim 26$ Myr
for the same depletion factor.
A 0.015 $\msol$ brown dwarf requires 50 Myr in the standard case and
70 Myr if [$^2D_0$] is twice as large to reach the  99\% depletion limit.
Conversely, if [$^2D_0$] is smaller by a factor of 2, the 99\% depletion limit
is reached in 1.7, 12 and 40 Myr for respectively 0.07, 0.02
and 0.015 $\msol$. 
If the initial [$^2D_0$] is as small as 2$\times$10$^{-6}$, {\it e.g} a factor of
10 smaller than the standard LISM value, all objects with masses
$m \, \simgr 0.04 \, \msol$  depletes their initial deuterium within 2 Myr.

The effect of a variation of [$^2D_0$] on isochrones
in CMDs, however, remains small, compared
to other sources of uncertainties at young ages (extinction,
$\te$ calibration, initial models) and
can be ignored for this present. If the improvement
of observable techniques in the future, however, allows the detection of
 deuterium in the atmosphere of young objects (Chabrier et al. 2000b), 
this effect needs to be taken into account for a correct age estimate
of very young clusters.

\subsection{Convection}

One of the major uncertainties affecting the evolution
of stars with masses $m \simgr 0.6 \msol$ is
due  to the treatment of convection. 
These stars show relatively extended
super-adiabatic outer layers during PMS and MS evolution,
of which description is extremely  sensitive to the adopted treatment of
convection, i.e., within the MLT formalism, to the adopted
mixing length (at any
gravity). The effect of a variation
of $l_{\rm mix}$ on evolutionary tracks for solar-type stars
is well known and is illustrated {\it e.g} in Figure 2b of Baraffe
et al. (2001). 
Ludwig, Freytag \& Steffen (1999)
calibrated the mixing length parameter $\alpha_{\rm mix}$ for these stars
with 
2D hydrodynamical models performed 
in the parameter space 4300 K $\le \te \le$ 7100 K and gravities
2.54 $\le \log g \le$ 4.74. They found a moderate variation of the
mixing length parameter around typically $\alpha=$1.5. 
  Note that for the present models  and input physics
\footnote{As
stressed in BCAH98, the value of $l_{\rm mix}$ required to fit the Sun
depends on the input physics (equation of state, opacities, outer boundary condition)},
 $l_{\rm mix}$ = 1.9 $H_{\rm P}$
is the value required to fit the Sun at its present age.
An increase of $\alpha_{\rm mix}$ from 1 to 2 yields an increase
of $\te$ up to 500K for the highest masses during their
PMS evolution.

For masses $m \simle 0.6 \msol$, the extension of the super-adiabatic
layers retracts appreciably and the transition from  convective to
radiative outer layers
is characterized by an abrupt transition from a fully adiabatic to
a radiative structure with a very small entropy jump. This means
that during most of the evolution, except at early ages, as discussed in \S 3.1.1 and
below,  
the sensitivity of the evolutionary  models to $l_{\rm mix}$ is
small. 
Multi-dimensional hydrodynamical simulations for conditions characteristic of M-dwarf atmospheres,
 $\te \, \le$ 3000 K, $\log g=5$, have been recently conducted
by Ludwig, Allard \& Hauschildt (2001). These simulations confirm the
afore-mentioned small entropy jump found in the 1-D models described by the MLT,
illustrating the large efficiency of atmospheric convection
 for these objects, a direct consequence of the formation of
molecules, as mentioned earlier. 
Under these circumstances, the 3D simulations show that the MLT does indeed provide a correct thermal profile, providing
a value of $\alpha_{\rm mix} \, > \,$ 1, at least for high gravities and
relatively old objects ($t >>$ 10 Myr). 

In \S 3.1.1,  however, we have  shown that even below 0.6 $\msol$,
 very young models with gravities 
$\log g \, \simle \,$ 4 can be affected by a variation of $l_{\rm mix}$. 
To minimize such uncertainties, 
a correct calibration of $l_{\rm mix}$ requires the extension of
the Ludwig et al. (2001) calculations to lower gravities.
This work is under progress and represents the most promising
method for an accurate description of convection in optically-thin media, 
through an accurate
calibration of $l_{\rm mix}$. In contrast, given all the already mentioned uncertainties
inherent to either observation or models 
for very young objects, a calibration of $l_{\rm mix}$ based on a comparison 
of PMS tracks with observations is at best highly speculative.
 
\section{Comparison with other work}

\begin{figure}
\psfig{file=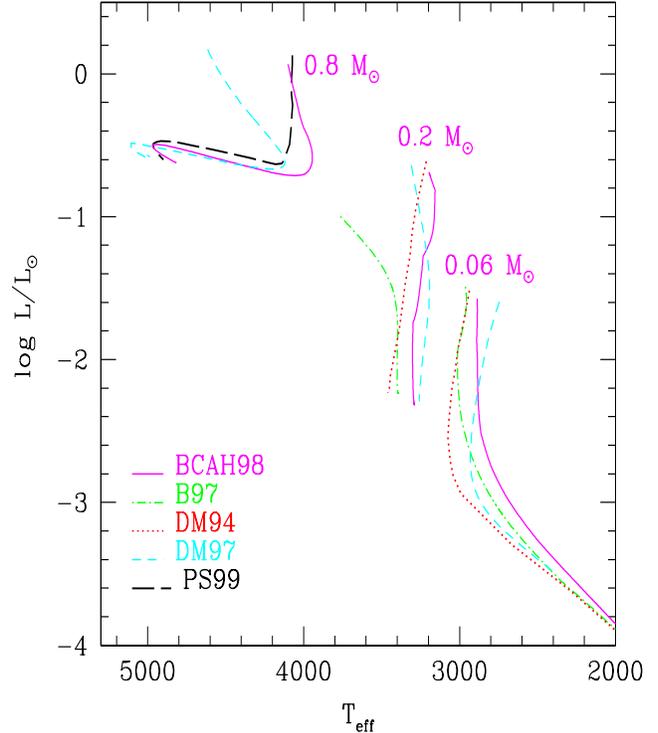,height=110mm,width=88mm} 
\caption{Comparison of evolutionary tracks from different authors, as
indicated on the figure. All models start at $t$ = 1 Myr, and
evolve up to the MS for the 0.8 and 0.2 $\msol$ stars.
}
\label{fig8}
\end{figure}

A detailed comparison between various
 evolutionary tracks
 available in the literature has already been done
by Siess, Dufour \& Forestini (2000). In the present section, we compare the 
evolutionary tracks  
most widely used by the community 
to describe the observational properties of objects in young clusters
(Burrows et al. 1997, B97; D'Antona \& Mazzitelli 1994, 1997, DM94, DM97; 
Palla \& Stahler 1999, PS99). 
The comparison between these different models in a theoretical HR diagram
is illustrated in Figure \ref{fig8} for a few low-mass star and BD masses.

Let us first summarize the main differences between these models. 
One crucial difference is due to the outer boundary condition:
DM94, DM97 and PS99 use approximate boundary conditions based on
$T(\tau)$ relationships assuming gray  approximation and 
radiative equilibrium.
Outer boundary conditions based on such approximations
are wrong as soon as 
molecules form near the photosphere and convection
reaches optically thin layers, {\it i.e} below $\te \sim 4000K$
(see Chabrier and Baraffe 1997, 2000; and references therein)
They usually yield hotter effective temperatures for a given mass.
Above $\sim 4000K$, the choice of the outer boundary condition
is less consequential, as shown by the good agreement between the 0.8 $\msol$ 
tracks of BCAH98 and of PS99 (see Figure \ref{fig8}),
all the other physical inputs being similar for such mass. 
Note that, although the PS99 models start from a more realistic birthline,
this does not affect the results after 1 Myr,
as expected from the tests on initial conditions
performed in \S 3.1.1.  

The B97 models for low mass stars and  hot brown dwarfs 
($\te \simgr 2000K$) are based on gray atmosphere
 models obtained by solving the radiative transfer
equation, as described in Burrows et al. (1993). Such an approximation,
although it represents an improvement over the previous 
$T(\tau)$ relationships, still overestimates $\te$ at a given mass
compared to models based on full non-gray atmosphere models.
This is certainly the reason why for VLMS on the Main Sequence
 (for $m > 0.08 \msol$), 
the B97 models are
about 100-200K hotter than the BCAH98 ones.
We note, however, that the very early evolution of the B97 models
proceeds at much hotter $\te$ than any other model. 
 Although also true for masses below 0.2 $\msol$, this does not appear
in Fig. \ref{fig8} since the tracks start at 1 Myr
and the 0.06 $\msol$ of B97 displayed in Fig. \ref{fig8}
is indeed much hotter but for $t < $ 1 Myr.
The reasons for
such differences along the Hayashi line
is not clear. We only notice that similar shape of the
Hayashi line ({\it i.e.} a strong decrease of $\te$) is obtained 
if one assumes a fully adiabatic initial structure and adiabatic convection 
({\it i.e}  $\alpha_{\rm mix} \to \infty$). Based on test cases, we find that 
such initial models are much hotter for a given $L$. If $\te$
is high enough for H$_2$ formation to be negligible, such models
start to evolve toward cooler $\te$.  Once the fraction of
H$_2$ becomes significant and the adiabatic gradient small enough near the
photosphere, evolution proceeds  almost vertically in the HRD, 
as also expected
from the analysis of Hayashi \& Nakano (1963) (see \S 3.1.1). 
Such an evolution resembles the shape of the B97 tracks.  It may however  be
a pure coincidence, since B97 do not describe their initial conditions.
 
Convection in all afore-mentioned models is treated within the framework of
 the mixing length theory,
with $1 < \alpha_{\rm mix} < 2$, except DM97, who use the Canuto-Mazzitelli
formalism. As demonstrated in \S 3, the treatment of convection affects
significantly evolutionary tracks and a variation of 
 $\alpha_{\rm mix}$ modifies the shape of the Hayashi lines
(as long as super-adiabatic layers are present). 
Figure \ref{fig8}
displays quite similar shapes for a given mass between BCAH98, PS99 and
DM94, although not at the same  $\te$  because of different outer boundary
conditions. The DM97 Hayashi lines behave differently with respect to
models of other groups, as already noticed by Siess et al. (2000). This is
certainly due to their different treatment of convection. 
Unexpectedly, the DM97 models predict a MS for VLMS close to the BCAH98 MS,
which, as mentioned by DM97, is purely coincidental and stems from 
unexpected canceling effects between different
 treatments of convection  and outer boundary conditions.
This is illustrated for the 0.2 $\msol$ in Figure \ref{fig8}.
Since the grey-like outer boundary condition used in DM97 is expected to yield hotter
effective temperatures than the BCAH98 models, the close agreement on the MS
suggests that the Canuto-Mazzitelli convection
treatment  yields larger super-adiabatic layers in the atmosphere.
This is indeed required to decrease $\te$ for a given mass.
Such a behaviour, however, is not found by the recent simulations by Ludwig et al. 
(2001) (see \S 3.2). Indeed, the Canuto-Mazzitelli treatment of convection yields results at odds with 3D hydro simulations
for the outer thermal profile of the Sun (see e.g. Nordlund \& Stein 1999), 
and does not provide an accurate treatment of convection
in optically-thin media, at least for solar-type stars and low-mass stars ( see e.g. \S 2.1.3  of Chabrier \& Baraffe, 2000 for a
discussion of this topic). 

\section{Observational tests}

In the previous sections, we emphasize the large uncertainty of
evolutionary models at early ages and low gravities. 
Testing different initial conditions, we find out 
that after a few Myr these uncertainties become inconsequential and
all our models converge toward the same position in a HRD at a given age, for a given mass.
Such a result
does not necessarily imply that the models are reliable at ages of a few Myr,
since we have only tested  simple cases. More sophisticated
initial conditions are beyond the scope of the present paper and
were already explored by Hartman et al. (1997), illustrating the
sensitivity of tracks and birthline positions
to the (poorly known) details of the protostellar
accretion process (geometry,
rate, temperature of added matter).  

A better knowledge of initial conditions may come from
the determination of the minimum
age below which present models start to depart
significantly from observations. 
Estimation of this age can constrain 
the characteristic time-scales and accretion rates
of the protostellar collapse phase.
Unfortunately, 
direct comparisons of observations with models directly in colour - magnitude diagrams
are extremely uncertain due to the large extinction in star formation regions, which affects the
observed energy distribution and thus the spectra and the colors. 
Only very few exceptions, such as
$\sigma$ Orionis,  exhibit low extinction. 
Recently, B\'ejar et al. (1999) and Zapatero et al. (1999, 2000) 
obtained optical and near-IR photometry
for low mass objects in this cluster. 
In a $(I-J)$ vs $M_{\rm I}$ CMD,
the data lie between the 1 and 10 Myr isochrones, respectively, 
for masses down to $\sim 0.01 \msol$,
using the BCAH98 
and CBAH00 models (Zapatero et al. 2000; B\'ejar et al. 2001). 
If  statistics is improved and if the
membership of the objects to the cluster is confirmed, such observations
provide an unique opportunity 
to test directly the validity of young theoretical isochrones. They also offer the best chance
to determine the Initial Mass Function (IMF) down to the substellar regime
and 
the minimum mass formed by a collapse process (see B\'ejar et al. 2001). 
 
\begin{figure*}
\psfig{file=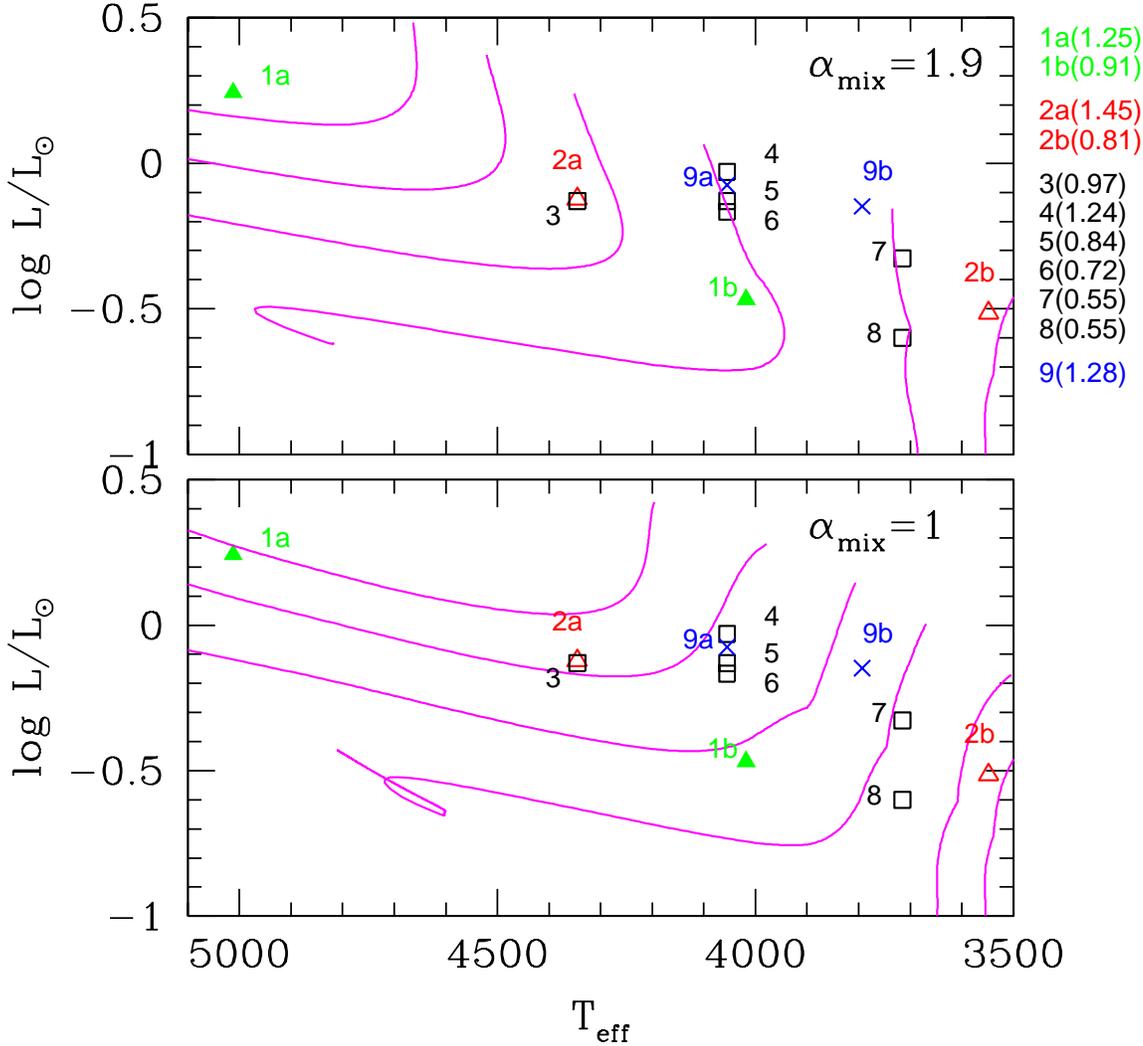,height=160mm,width=160mm} 
\caption{Comparison of evolutionary tracks with observed PMS  
objects with derived masses. The BCAH98 tracks are displayed
for 1.4, 1.2, 1, 0.8, 0.6 and 0.5 $\msol$ (from left to right) for
two values of $\alpha_{\rm mix}$. Observations are from 
Covino et al. (2000, filled triangles), Simon et al. (2000, crosses and
open squares) and  
Steffen et al. (2001, open triangles). Open squares are single objects  whereas
all the other symbols indicate binaries. For each systems, masses are indicated
(in $\msol$) on the right hand side of the figure. 
For GG tau A (9a+9b, crosses) the total mass is indicated.
}
\label{fig9}
\end{figure*}

Young multiple systems provide also excellent tests for
PMS models at young ages, because of the assumed coevality of
their different components. In addition, another strong constraint 
is supplied by the estimate of
dynamical masses deduced either from binary systems (Covino et al. 2000; Steffen et al. 2001)
or determined from
the orbital motion of 
circumstellar/circumbinary disks (Simon, Dutrey \&  Guilloteau 2000). 
An example is provided by the quadruple system GG TAU (White et al. 1999),
with components covering the whole mass-range of VLMS and BDs from 1 $\msol$
to $\sim$ 0.02 $\msol$. Orbital velocity measurements of the circumbinary
disk surrounding the two most massive components imply a constraint
on their combined stellar mass (Dutrey, Guilloteau \& Simon 1994;
 Guilloteau, Dutrey \& Simon 1999). This mass constraint and the hypothesis
of coevality provides a stringent test for PMS models. 
The BCAH98 models   
are the most consistent with GG Tau 
(for details see White et al 1999;  Luhman 1999) and  
provide the closest agreement with derived masses of other young systems (see
Fig.\ref{fig9}). 

Most of the observed
systems displayed in Figure \ref{fig9} are better reproduced by tracks
using a large
value of $\alpha_{\rm mix}$ (=1.9).  
However, for some systems, such as 1 
(Covino et al. 2000), 2 (Steffen et al. 2001)  and 4 (BP Tau from Simon
et al. 2000), a better agreement is
obtained with  $\alpha_{\rm mix}$ = 1. 
Although a variation of $\alpha_{\rm mix}$
with effective temperature and gravity is possible, as suggested
by the simulations of Ludwig et al. (1999), none of these three systems
occupies a peculiar position in ($\te$, $g$) to
suggest a different value of $\alpha_{\rm mix}$. 
This puzzle may reflect the uncertainties of PMS models based on 
arbitrary initial conditions. It may also be due to
the large uncertainties of observationally-derived
spectral type classifications, luminosity estimates and  
$\te$ calibrations for such very young objects. 
Since they  display spectral
features between that of giants and dwarfs (see Luhman 1999), a better representation
of their spectral properties may require new indices 
more appropriate to these intermediate surface
gravities, in the same vein as the pseudo-continuum ratios used by
Mart\'\i n et al. (1996)  for Pleiades objects. 
The transformation of the inferred spectral type into $\te$
is even more difficult, because of the lack of reliable 
$\te$ - scales for such young T-Tauri like objects. Significant
efforts were devoted  within the past recent years to the elaboration
of improved $\te$ - scales for M-dwarfs (Leggett et al. 1996) 
and M-giants (Perrin et al. 1998; van Belle et al. 1999). However, work
remains to be done for T-Tauri like objects. 
Luhman (1999) defined a $\te$ - scale intermediate between giants
and dwarfs and based on the isochrone of BCAH98 which
goes through the four components of GG Tau. Interestingly enough, applying
this $\te$ - scale to young clusters such as IC348 (Luhman 1999) and
star forming regions like Chamaeleon I (Comer\'on, Neuh\"auser \& Kaas 2000),
the cluster members show a small scatter in age and no obvious
correlation between age and mass. As mentioned by Comer\'on
et al. (2000), this suggests in Chamaeleon I an almost
coeval population which formed within less than 1 Myr.
This short timescale supports the suggestion that star formation is
 controlled primarily by large-scale turbulent flows rather than by magnetic 
processes such as ambipolar diffusion (Hartmann 2000).

Given the assumption of coevality and the uncertainties of
our initial conditions, the $\te$ - scale for young objects suggested 
by Luhman (1999)
needs to be confirmed  in order to confirm these exciting results about star formation process.
Although still very preliminary,
the comparison of observed and synthetic
spectra, as recently done by Lucas et al. (2001) for Orion objects, provides
a promising way to define such a $\te$ - scale.  

\section{Conclusion}

The very good agreement of models based on improved physics with observations
for relatively old (t $\simgr$ 100 Myr) low-mass objects yields
confidence in the underlying theory. Such evolutionary models
can now be confronted to the complex realm of very young objects,
providing important information on star formation processes and 
initial conditions for PMS models. 
Although based on extremely simple initial conditions (no accretion phase,
no account of protostellar collapse phase and time scale, 
spherical symmetry),
these models provide the most accurate comparison with present observations of
very young objects
(dynamical masses, tests of coevality in multiple systems, CMDs).
Given the combining effects of large observational and theoretical uncertainties
at very young ages, however, one must remain cautious. It is probably too premature to conclude
on the validity of the present models at early phases of evolution.
We have examined in the present paper the uncertainties on evolutionary models of very-young
low-mass objects arising from initial conditions, in particular the initial radius of the object, the efficiency of
convection in the outermost layers and the
initial abundance of deuterium. We have shown that at least the two first afore-mentioned uncertainties
can affect drastically the fundamental properties, luminosity and effective temperature of objects younger
than about 1 Myr. Therefore, any attempt to infer an age or a mass from observable quantities
for these objects, in particular the initial mass function of very young clusters, must be considered with highly
limited - if any - validity !

Realistic initial conditions can only be provided
by multi-dimensional protostar collapse simulations, not by spherically-symmetric models
for PMS initial conditions 
involving many free, ill- or unconstrained parameters.
Because of numerical subtleties and complex physical
processes (accretion fronts, turbulent time-dependent convection, hydrodynamical
radiative transfer, magnetic field etc...), the construction of star formation
models is a harsh  task, which very likely will necessitate several years of efforts.
Besides these theoretical difficulties, observations of very young objects can provide only limited guidance
to such simulations, since most phases involved during
the collapse are embedded in dusty cocoons. Only the final product
can be observationally tested, when the protostar becomes visible.
This stage marks essentially the beginning of PMS evolutionary tracks. 
PMS tracks tested against observations thus provide a precious link to gather insight about
star formation models from subsequent evolution. 

To progress in the field and in parallel with the development
of star formation models, efforts can  be directed toward:
(1) a reduction of the main theoretical uncertainties affecting PMS models
at low gravities, which involves in particular a better determination of $l_{\rm mix}$
through multi-D hydrodynamical simulations and (2)
the elaboration of a reliable $\te$-scale for young objects.

\medskip
{\it Note: }
Tracks and isochrones for $t \, \ge$ 1 Myr of the BCAH98 models
(from 0.02 $\msol$ to 1.4 $\msol$)  and of the CBAH00 models including
dusty atmospheres (from 0.001 $\msol$ to 0.1 $\msol$)
are available by anonymous ftp:
\par
\hskip 1cm ftp ftp.ens-lyon.fr \par
\hskip 1cm username: anonymous \par
\hskip 1cm ftp $>$ cd /pub/users/CRAL/ibaraffe \par
\hskip 1cm ftp $>$ get README \par
\hskip 1cm ftp $>$ get BCAH98\_models.* \par
\hskip 1cm ftp $>$ get BCAH98\_iso.* \par
\hskip 1cm ftp $>$ get DUSTY00\_models \par
\hskip 1cm ftp $>$ quit
\bigskip

\begin{acknowledgements} We are indebted to Lee Hartmann for valuable 
discussions. I.B thanks the Max-Planck Institut for Astrophysik in Garching
for hospitality during elaboration of part of this work.
The calculations were performed using facilities at Centre
d'Etudes Nucl\'eaires de Grenoble.
\end{acknowledgements}


\begin{thebibliography}{}

\bibitem[]{} Allard, F., Hauschildt, P. H., Alexander, D. R., \& Starrfield, S.
 1997, \araa, 35, 137
\bibitem[]{} Allard F., Hauschildt P.H., Schweitzer, A., 2000, \apj, 539, 366
\bibitem[]{} Allard, F., Hauschildt, P.H., Alexander, D.R., Tamanai, A.,
Schweitzer, A.. 2001, \apj, 556, 357
\bibitem[]{} Baraffe I., Chabrier G., Allard F., Hauschildt P.H., 1995, \apjl,
446, 35
\bibitem[]{} Baraffe I., Chabrier G., Allard F., Hauschildt P.H., 1997, A\&A 327, 1054
\bibitem[]{} Baraffe I., Chabrier G., Allard F., Hauschildt P.H. 1998, \aap, 337, 403 (BCAH98)
\bibitem[]{} Baraffe I., Chabrier G., Allard F., Hauschildt P.H. 2001,
 {\it From darkness to light: origin and evolution of young
stellar clusters}, ASP Conf Series, Vol. 243, Cargese 2000, astro-ph/0007157
\bibitem[]{} B\'ejar, V.J.S., Zapatero Osorio M.R., Rebolo R. 1999, 
\apj, 521, 671
\bibitem[]{} B\'ejar, V.J.S., Mart\'\i n, E.L., Zapatero Osorio, M.R., 
Rebolo, R.,
Barrado y Navascu\'es, D., Bailer-Jones, C.A.L., Mundt, R., 
Baraffe, I., Chabrier, G., Allard, F. 2001, \apj, 556, 830
\bibitem[]{} Burrows A., Liebert, J. 1993, Rev. Mod. Phys., 65, 301
\bibitem[]{} Burrows A., Hubbard, W.B., Saumon, D., and Lunine, J. I., 1993,
 \apj, 406, 158
\bibitem[]{} Burrows A., Marley M., Hubbard W.B., 
Lunine, J.I., Guillot, T., Saumon, D.,
Freedman, R., Sudarsky, D., Sharp, C. 1997, \apj, 491, 856
(B97)
\bibitem[]{} Chabrier, G., Baraffe, I. 1997, \aap, 327, 1039
\bibitem[]{} Chabrier, G., Baraffe, I. 2000, \araa, 38, 337 
\bibitem[]{} Chabrier, G., Baraffe, I., Allard, F., Hauschildt, P.H. 2000a, 
\apj, 542, 464 (CBAH00)
\bibitem[]{} Chabrier, G., Baraffe, I., Allard, F., Hauschildt, P.H. 2000b, 
\apjl, 542, 119
\bibitem[]{}Comer\'on, F., Neuh\"auser, R., \& Kaas, A.A. 2000, 
\aap, 359, 269
\bibitem[]{} Covino, E., Catalano, S., Frasca, A., Marilli, E., Fernandez, M., Alcala, J.M., Melo, C., Paladino, R., Sterzik, M.F., Stelzer, B. 2000, \aap,
361, L49
\bibitem{} D'Antona, F. and Mazzitelli, I, 1994, \apjs, 90, 467 (DM94)
\bibitem{} D'Antona, F. and Mazzitelli, I, 1997, in ``Cool stars 
in Clusters and Associations'', eds R. Pallavicini and G. Micela,
 Mem. S. A. It., 68, 807 (DM97)
\bibitem[]{} Dutrey, A., Guilloteau, S., \& Simon, M. 1994, \aap, 286, 149
\bibitem[]{} Guilloteau, S., Dutrey, A.,  \& Simon, M. 1999, \aap, 348, 570
\bibitem[]{} Hartmann, L. 2000, in Proceedings of 33rd ESLAB Symposium, 
"Star formation from the small to
 the large scale", F. Favata, A.A. Kaas, and A. Wilson eds., ESA SP-445
\bibitem[]{} Hartmann, L., Cassen, P., Kenyon, S.J. 1997, \apj, 475, 770
\bibitem[]{} Hauschildt P.H., Allard F., Baron E. 1999a, \apj, 512, 377
\bibitem[]{} Hauschildt P.H., Allard F., Ferguson, J., Baron, E., 
Alexander, D.R. 1999b, \apj, 525, 871
\bibitem[]{} Hayashi, C. 1961, PASJ, 13, 450
\bibitem[]{} Hayashi, C., Nakano, T. 1963, Prog. Theor. Phys., 30, 460
\bibitem[]{} Leggett, S.K., Allard, F., Berriman, G., Dahn, C.C.,  \&
Hauschildt, P.H. 1996, \apjs, 104, 117
\bibitem []{} Linsky, J. 1998, Space Science Reviews, 84, 285
\bibitem []{} Lucas, P.W., Roche, P.F., Allard, F., Hauschildt, P.H. 2001,
\mnras, 326, 695
\bibitem[]{} Ludwig, H.G., Freytag, B., \& Steffen, M. 1999, \aap, 346, 111
\bibitem[]{} Ludwig, H.G., Allard, F., Hauschildt, P.H. 2001, \aap, submitted
\bibitem[]{} Luhman KL. 1999,  \apj,  525, 466
\bibitem[]{} Mart\'\i n, E.L., Rebolo, R., Zapatero Osorio, M.R. 1996, 
\apj, 469, 706
\bibitem[]{} Miller S., Tennyson J., Jones H.R.A, Longmor, A.J., 1994,
in Molecules in the Stellar Environment, ed. U.G Jorgensen, Lecture Notes
in Physics
\bibitem[]{} Nordlund, A., Stein, R.F. 1999,
``Theory and Tests of Convection
 in Stellar Structure'', ASP Conf. Series 173, p.91
\bibitem[]{} O'Meara, J.M., Tytler, D., Kirkman, D., Suzuki, N., 
Prochaska, J.X., Lubin, D., Wolfe, A.M.
2001, AAS, 197, 5604 (astro-ph/0011179)
 \bibitem[]{}Palla, F., \& Stahler, S.W. 1999, \apj, 525, 772 (PS99)
\bibitem[]{} Partridge, H., and Schwenke, D.W., 1997, {\it J. Chem. Phys.}, 106, 4618
\bibitem[]{} Perrin, G., Coud\'e du Foresto, 
V., Rigway, S.T., Mariotti, J.-M., 
Traub, W.A., Carleton, N.P., \& Lacasse, M.G. 1998, \aap, 331, 619
\bibitem[]{} Saumon D., Hubbard W.B., Burrows A., Guillot T., Lunine J.I.,
Chabrier, G. 1996, \apj, 460, 993
\bibitem[]{} Schwenke, D.W., 1998, {\it Chemistry and Physics of Molecules and Grains in Space}, Faraday Discussion 109, 321
\bibitem[]{} Siess, L., Dufour, E., \& Forestini, M. 2000, \aap, 358, 593
\bibitem[]{} Simon, M., Dutrey, A., \& Guilloteau, S. 2000, \apj, 545, 1034
\bibitem[]{} Stahler, S.W., 1983, \apj, 274, 822
\bibitem[]{} Stahler, S.W., 1988, \apj, 332, 804
\bibitem[]{} Stahler, S.W., Shu, F.H., Taam, R.E. 1980, \apj, 241, 637
\bibitem[]{} Steffen, A.T., Mathieu, R.D., Lattanzi, M.G., Latham,
D.W., Mazeh, T., Prato, L., Simon, M., Zinnecker, H., Loreggia,
D. 2001, \aj, 122, 997
\bibitem[]{} van Belle, G.T., et al. 1999, \apj, 117, 521
\bibitem[]{} White R.J., Ghez A.M., Reid I.N., Schultz G. 1999, \apj, 520, 811
\bibitem[]{} Zapatero Osorio, M.R., B\'ejar V.J.S., Rebolo, R.,
 Mart\'\i n, E.L., Basri, G. 1999, \apjl, 524, 115
\bibitem[]{} Zapatero Osorio, M.R., B\'ejar V.J.S., Mart\'\i n, E.L., 
Rebolo, R., Barrado y Navascu\'es, D., Bailer-Jones, C.A.L., 
Mundt, R. 2000,  Sciences, 290, 103

\end{thebibliography}
\end{document}